\documentclass[%twocolumn
]{jpsj3}
\usepackage{txfonts}
\usepackage{color}

\title{Correlation Effects on Charge Order and Zero-Gap State in the Organic Conductor 
$\alpha$-(BEDT-TTF)$_2$I$_3$}

\author{Yasuhiro Tanaka$^{1}$ and Masao Ogata$^{2}$}
\inst{$^{1}$Department of Physics, Chuo University, Bunkyo, Tokyo 112-8551, Japan \\
$^{2}$Department of Physics, University of Tokyo, Bunkyo, Tokyo 113-0033, Japan} %\\

\abst{
The effects of electron correlation in the quasi-two-dimensional organic conductor 
 $\alpha$-(BEDT-TTF)$_2$I$_3$ are investigated theoretically by using an extended Hubbard model 
with on-site and nearest-neighbor Coulomb interactions. 
A variational Monte Carlo method is applied to study its ground-state properties. 
We show that there appears a nonmagnetic 
horizontal-stripe charge order in which nearest-neighbor correlation functions indicate a 
tendency toward a spin-singlet formation on the bonds with large transfer integrals along the 
charge-rich stripe. 
Under uniaxial pressure, a first-order transition from the nonmagnetic charge 
order to a zero-gap state occurs. 
Our results on a spin correlation length in the charge-ordered state suggest that 
a spin gap is almost unaffected by the uniaxial pressure in spite of the suppression of the charge 
disproportionation. The relevance of these contrasting behaviors in spin and charge 
degrees of freedom to recent experimental observations is discussed.}
 
\begin{document}
\maketitle

\section{Introduction}
Low-dimensional organic conductors have been studied intensively\cite{Ishiguro_BOOK}. 
They exhibit intriguing electronic phases that originate from a variety of their crystal 
structures and strong Coulomb interactions among 
electrons.\cite{Seo_Rev} Typical examples are Mott insulators and 
charge-ordered states in which the effects of electron correlation play an important 
role in understanding their physical properties.  

Among various compounds, the quasi-two-dimensional organic conductor 
$\alpha$-(BEDT-TTF)$_2$I$_3$ [BEDT-TTF=bis(ethylenedithio)tetrathiafulvalene], which we 
abbreviate as $\alpha$-I$_3$ hereafter, has attracted much attention\cite{Kajita_Rev}. 
It shows a metal-insulator transition at $T_{\rm CO}=135$ K\cite{Bender_MCLC84}, which is of 
first order with a small 
structural distortion\cite{Heidmann_SSC92,Emge_MCLC86,Fortune_SSC91}. 
The conduction layer of $\alpha$-I$_3$ consists of donor BEDT-TTF molecules and 
forms a 3/4-filled band since the anion $I^{-}_3$ is monovalent. Below $T_{\rm CO}$, 
a horizontal-stripe charge order (CO) appears\cite{Takano_JPCS01,Takano_SM01,Wojciechowski_PRB03,Kakiuchi_JPSJ07,Kawai_JPSJ09,Yue_PRB10} together with a spin-gap behavior as observed in the magnetic 
susceptibility measurement\cite{Rothamael_PRB86}. 
By applying uniaxial pressure, this CO changes into a zero-gap state (ZGS)\cite{Tajima_JPSJ06}. 
Various anomalous transport phenomena observed in this ZGS\cite{Kajita_Rev} are 
considered to originate from massless Dirac fermions\cite{Katayama_JPSJ06} 
located at the Fermi level in the energy dispersion. 

Theoretically, several authors have investigated the physical properties of $\alpha$-I$_3$ in its 
various electronic phases including 
the CO\cite{Kino_JPSJ95,Kino_JPSJ96,Seo_JPSJ00,Tanaka_JPSJ08,Miyashita_JPSJ08}, 
superconductivity\cite{Kobayashi_JPSJ04,Kobayashi_JPSJ05}, and the ZGS
\cite{Katayama_JPSJ06,Kobayashi_JPSJ07,Kobayashi_JPSJ13,Kobayashi_PRB11,Katayama_EPJ09,
Isobe_JPSJ12,Kobayashi_JPCS08,Kobayashi_Sci09,Geerbig_PRB08,Morinari_JPSJ14,Piechon_JPSJ13,Suzumura_JPSJ13b,
Monteverde_PRB13,Mori_JPSJ13,Piechon_JPSJ13b,Suzumura_JPSJ13,Nishine_JPSJ10,Kobayashi_JPSJ08,
Kobayashi_JPSJ09,Kino_JPSJ06,Suzumura_JPS11,Proskurin_JPSJ13,Suzumura_JPSJ14,Suzumura_JPSJ14_2}. 
By using an extended Hubbard model, Kobayashi {\it et al.} have found that the CO becomes 
unstable and the ZGS appears with increasing uniaxial pressure\cite{Kobayashi_JPSJ05}, 
the effect of which is incorporated in the modification of transfer integrals. 
They have also shown that the peculiar network of 
transfer integrals in this compound is crucially important for the realization of the Dirac-type 
energy dispersion\cite{Katayama_JPSJ06}. Although the above studies have explained 
some aspects of recent experimental findings\cite{Kajita_Rev}, our understanding on 
this system is still limited in the sense that most of these theories 
are based on the mean-field approximation (MFA) that ignores electron correlation effects. 
For example, the MFA predicts an antiferromagnetic spin order in the 
CO\cite{Seo_JPSJ00,Kobayashi_JPSJ05,Tanaka_JPSJ08,Kobayashi_PRB11}, 
which is inconsistent with experiments showing the nonmagnetic CO
%whereas $\alpha$-I$_3$ becomes nonmagnetic 
below $T_{\rm CO}$\cite{Rothamael_PRB86}. 
The importance of the correlation effects is also suggested by an NMR study for 
$T>T_{\rm CO}$\cite{Hirata_PRB11}. 
Recent experiments under hydrostatic pressure have shown that the pressure 
dependences of the CO and the spin gap are considerably different from 
each other\cite{Liu_DT16,Liu_PRL16}. Namely, the former is linearly suppressed until it 
disappears at around 10 kbar, 
whereas the latter is almost constant and steeply decreases near the phase boundary between 
the CO and the ZGS. Moreover, its effects on 
the massless Dirac fermions have not been fully explored yet. 

With these facts in mind, we investigate the ground-state properties of $\alpha$-I$_3$ by 
employing a variational Monte Carlo (VMC) method, which can take account of the quantum 
fluctuations. First, we show that a horizontal-stripe CO without the antiferromagnetic 
order is obtained for the first time, which is in contrast to the MFA. 
By increasing the uniaxial pressure, the nonmagnetic CO is suppressed and the ZGS becomes 
the ground state. In the CO state, nearest-neighbor spin-spin correlation functions 
indicate that a spin-singlet formation is favored on the 
bonds with large transfer integrals on the charge-rich stripe. This is because the 
magnitude of transfer integrals is alternating on the stripe, as pointed out by 
Seo\cite{Seo_JPSJ00} within the MFA. 
The results are also consistent with the exact diagonalization study on small clusters 
with electron-phonon interactions\cite{Miyashita_JPSJ08} that explicitly break the 
symmetry of the lattice. 
Furthermore, we show that the spin correlation length in the 
CO phase has very weak pressure dependence, which suggests the robustness of the spin gap. 
We discuss its origin and relevance to the experimental observations\cite{Liu_DT16,Liu_PRL16}.

\section{Model and Method}
\begin{figure}
\begin{center}
\includegraphics[height=5.0cm]{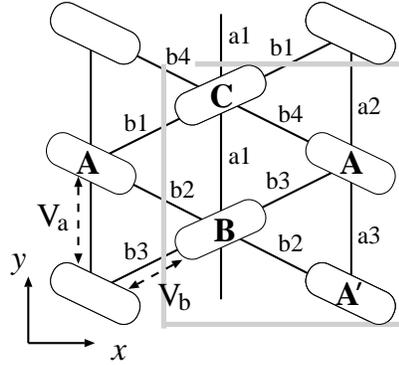}
\end{center}
\caption{Schematic representation of the crystal structure for $\alpha$-I$_3$ in the metallic phase. 
The gray rectangle indicates the unit cell.}
\label{fig:fig1}
\end{figure}

We consider the two-dimensional extended Hubbard model, which is written as
\begin{equation}
{\it H}=\sum_{\langle ij\rangle \sigma}t_{ij}(c^{\dagger}_{i\sigma}c_{j\sigma}+h.c.)
+U\sum_{i}n_{i\uparrow}n_{i\downarrow}+\sum_{\langle ij\rangle}V_{ij}n_in_j,
\label{eq:ham}
\end{equation}
where $\langle ij\rangle$ represents a pair of neighboring sites, $c^{\dagger}_{i\sigma}$ 
($c_{i\sigma}$) denotes the creation (annihilation) operator for an electron with spin 
$\sigma$ at the $i$th site, $n_{i\sigma}=c^{\dagger}_{i\sigma}c_{i\sigma}$, 
and $n_i=n_{i\uparrow}+n_{i\downarrow}$. 
We consider a 1/4-filled system, which is equivalent to the 3/4-filled band for 
$\alpha$-I$_3$ through a unitary transformation. 
The transfer integrals are written by $t_{ij}$, whereas 
$U$ and $V_{ij}$ indicate on-site and nearest-neighbor Coulomb interactions, respectively. 
The crystal structure of $\alpha$-I$_3$ in the high-temperature metallic phase is schematically 
shown in Fig. \ref{fig:fig1}. 
The unit cell contains four molecules, which we denote by A, A$^{\prime}$, B, and C. 
In the metallic phase, sites A and A$^{\prime}$ are equivalent to each other owing to the inversion 
symmetry, whereas they become inequivalent in the CO phase where sites A and B 
(A$^{\prime}$ and C) become charge-rich (charge-poor). Although the structural 
distortion slightly modulates the transfer integrals at $T_{\rm CO}$, we do not take its effects 
into account for simplicity\cite{Tanaka_JPSJ08,Miyashita_JPSJ08}. For $V_{ij}$, we consider $V_a$ and $V_b$ on 
the vertical and diagonal bonds, respectively, as shown in Fig. \ref{fig:fig1}. 

We write the trial wave function for the VMC calculations as
\begin{equation}
|\Psi \rangle = P_W P_G|\Phi \rangle ,
\label{eq:wf}
\end{equation}
where $|\Phi \rangle$ indicates the wave function obtained by diagonalizing 
the mean-field Hamiltonian
\begin{equation}
{\it H}^{\rm MF}=\sum_{\langle ij\rangle \sigma}
t_{ij}(c^{\dagger}_{i \sigma}c_{j \sigma}+h.c.)
+\sum_{\mu\alpha\sigma}\Delta_{\alpha\sigma}n_{\mu\alpha\sigma},
\label{eq:ham_mf}
\end{equation}
where $\mu$ denotes the index for the unit cell and $\alpha$ is 
that for the site in the unit cell. $P_G$ and $P_W$ are projection operators for 
on-site and nearest-neighbor sites, respectively\cite{Yokoyama_JPSJ90,Yokoyama_JPSJ04,Watanabe_JPSJ05,Watanabe_JPSJ06}. They are defined as
\begin{equation}
P_G=\prod_{i}[1-(1-g)n_{i\uparrow}n_{i\downarrow}],
\end{equation}
\begin{equation}
P_W=w_a^{\sum_{\langle ij\rangle_{a}}n_in_j}w_b^{\sum_{\langle ij\rangle_{b}}n_in_j},
\end{equation}
where $g\geq 0$, $w_a,\ w_b\leq 1$, and $\langle ij\rangle_{a}$ ($\langle ij\rangle_{b}$) stands for 
a pair of neighboring sites that are connected by the vertical (diagonal) bonds 
(see Fig. \ref{fig:fig1}). The variational parameters $\Delta_{\alpha \sigma}$, $g$, $w_a$, and $w_b$ 
are optimized by the stochastic reconfiguration method\cite{Sorella_PRB01}. Typically, physical 
quantities are measured using 
$2$--$6\times 10^5$ samples, which give sufficient statistical accuracy. 
In this paper, we do not consider any CO 
pattern with a unit cell which is different from that in Fig. \ref{fig:fig1}\cite{Tanaka_JPSJ08}. 
For the transfer integrals, we use $t_{b1}=-0.127$, $t_{b2}=-0.145$, $t_{b3}=-0.062$, 
$t_{b4}=-0.025$, $t_{a1}=0.035$, $t_{a2}=0.046$, and $t_{a3}=-0.018$\cite{Kakiuchi_JPSJ07}. 
The unit of energy is eV hereafter. We take account of the effect of uniaxial pressure $P$ (kbar) 
along the $y$-axis by modifying the transfer integrals $t_{ij}$ as $t_l(P)=t_l(1+K_lP)$ with 
$K_l$ (eV/kbar) given in Ref. 22. 

\section{Results}
\subsection{Effects of $P_G$}
\begin{figure}
\begin{center}
\includegraphics[height=4.5cm]{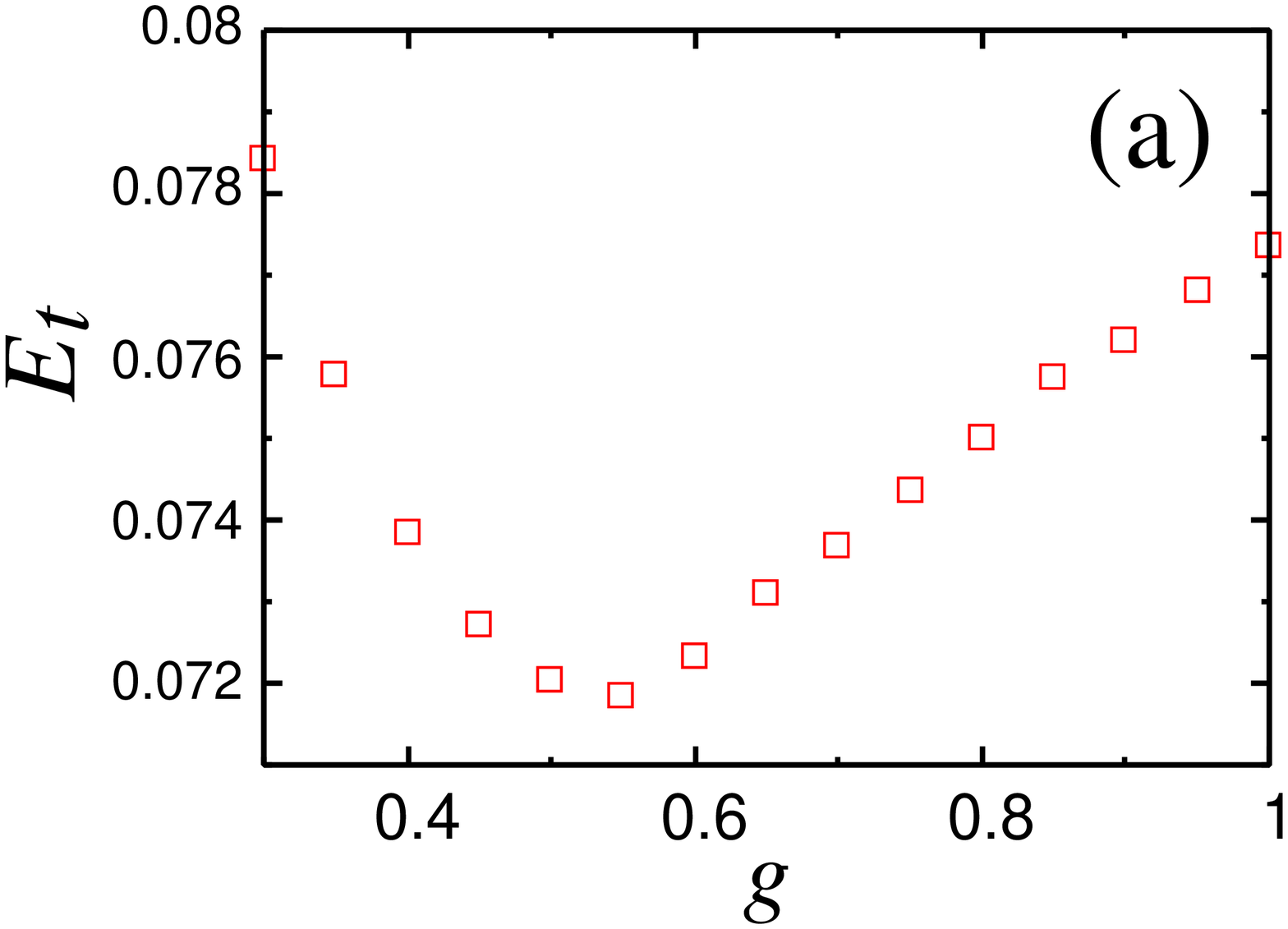}
\includegraphics[height=4.5cm]{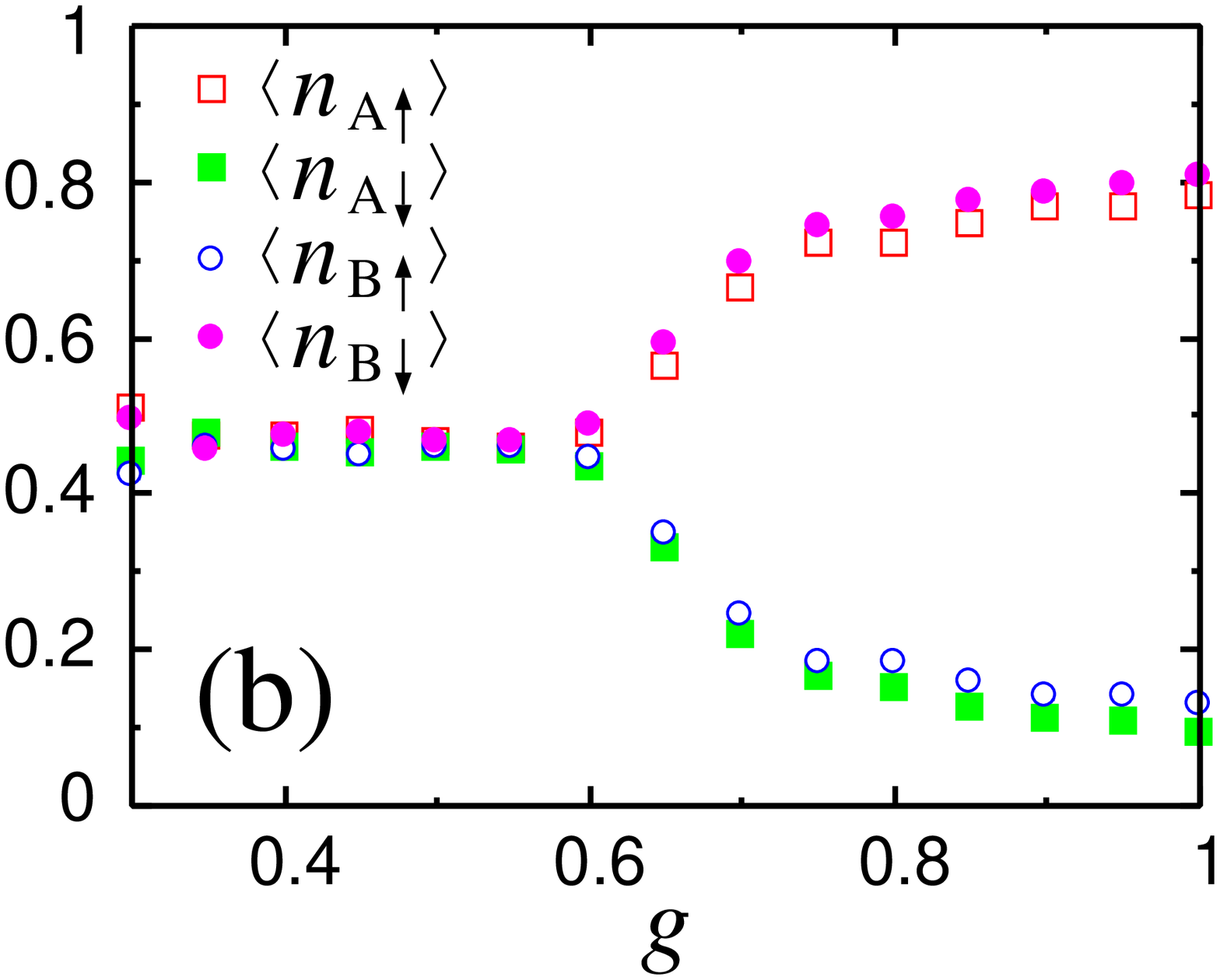}
\includegraphics[height=4.5cm]{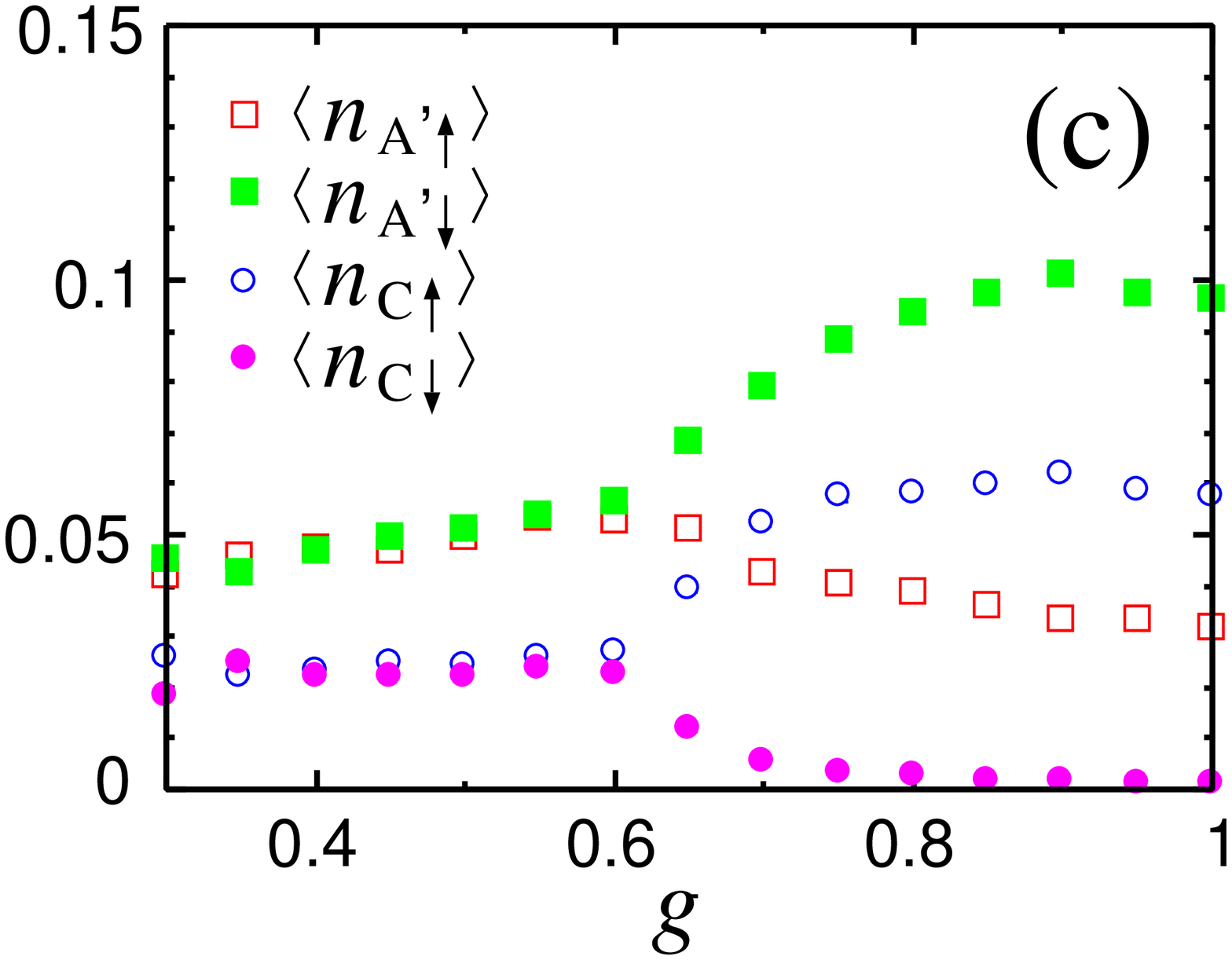}
\end{center}
\caption{(Color online) (a) Ground-state energy $E_t$, (b) average charge density 
$\langle n_{\alpha\sigma}\rangle$ for $\alpha=A, B$, and (c) those for $\alpha=A^{\prime}, C$ 
as a function of $g$ with $w_a=w_b=1$. We use $U=0.7$, $V_a=0.35$, $V_b=0.25$, and $P=0$ kbar.}
\label{fig:fig2}
\end{figure}

Before discussing the results with fully optimized wave functions, we first 
examine the effects of $P_G$ by setting $P_W=1$ in Eq. (2). 
The values of $\Delta_{\alpha \sigma}$ are optimized. We use 
$U=0.7$, $V_a=0.35$, and $V_b=0.25$ unless otherwise noted. 
Figure \ref{fig:fig2}(a) shows the ground-state energy $E_t$ as a function 
of $g$ for $P=0$ kbar. 
The system size is $N_u=L_x\times L_y=8\times 8$ with $L_x$ ($L_y$) being a 
number of unit cells in the $x$-direction ($y$-direction). We show the average charge density 
$\langle n_{\alpha\sigma}\rangle$ in the unit cell in Figs. \ref{fig:fig2}(b) and \ref{fig:fig2}(c). 
For $g=1$, the results are equivalent to those with the MFA\cite{Tanaka_JPSJ08}, 
which have CO together with antiferromagnetic order. 
We obtain a horizontal-stripe CO in which 
the sites $A$ and $B$ ($A^{\prime}$ and $C$) are charge-rich (charge-poor). In this state, 
we have $\langle n_{A\uparrow}\rangle > \langle n_{A\downarrow}\rangle$ and 
$\langle n_{B\downarrow}\rangle > \langle n_{B\uparrow}\rangle$ so that the spins 
on the charge-rich stripe form an antiferromagnetic order. As $g$ decreases, 
$E_t$ becomes lower. Correspondingly, the spin order is suppressed 
and it eventually disappears at $g=0.6$. Since the lowest $E_t$ is located at $g=0.55$, 
the ground state is a nonmagnetic CO. This shows a qualitative 
difference with the MFA\cite{Seo_JPSJ00,Kobayashi_JPSJ05,Tanaka_JPSJ08,Kobayashi_PRB11}.
Similarly, on the charge-poor sites ($A^{\prime}$ and $C$), a weak spin 
order is obtained for $g=1$, which is also suppressed with decreasing $g$. 
The origin of this nonmagnetic state will be discussed in Sect. 3.3. 

\begin{figure}
\begin{center}
\includegraphics[height=4.5cm]{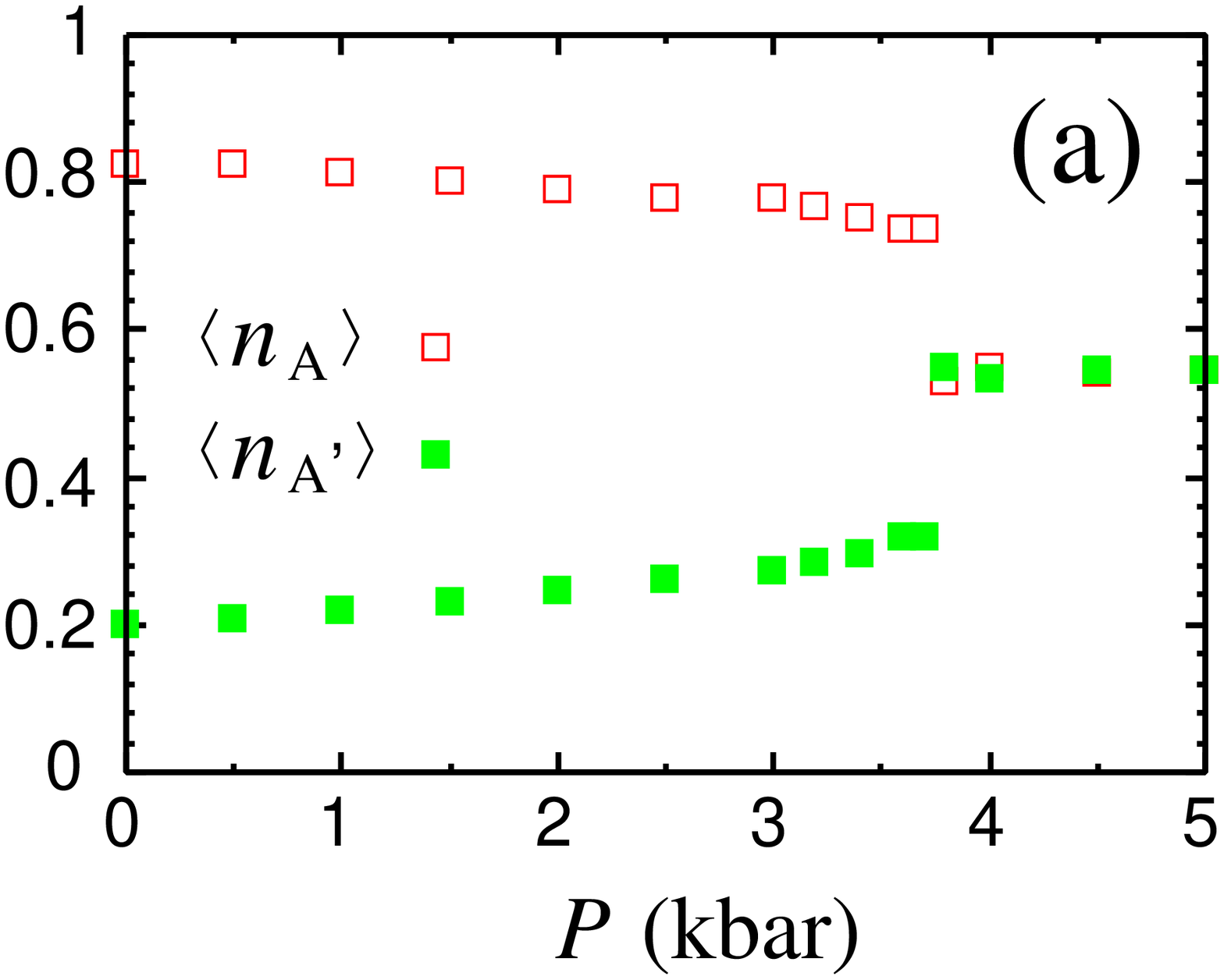}
\includegraphics[height=4.5cm]{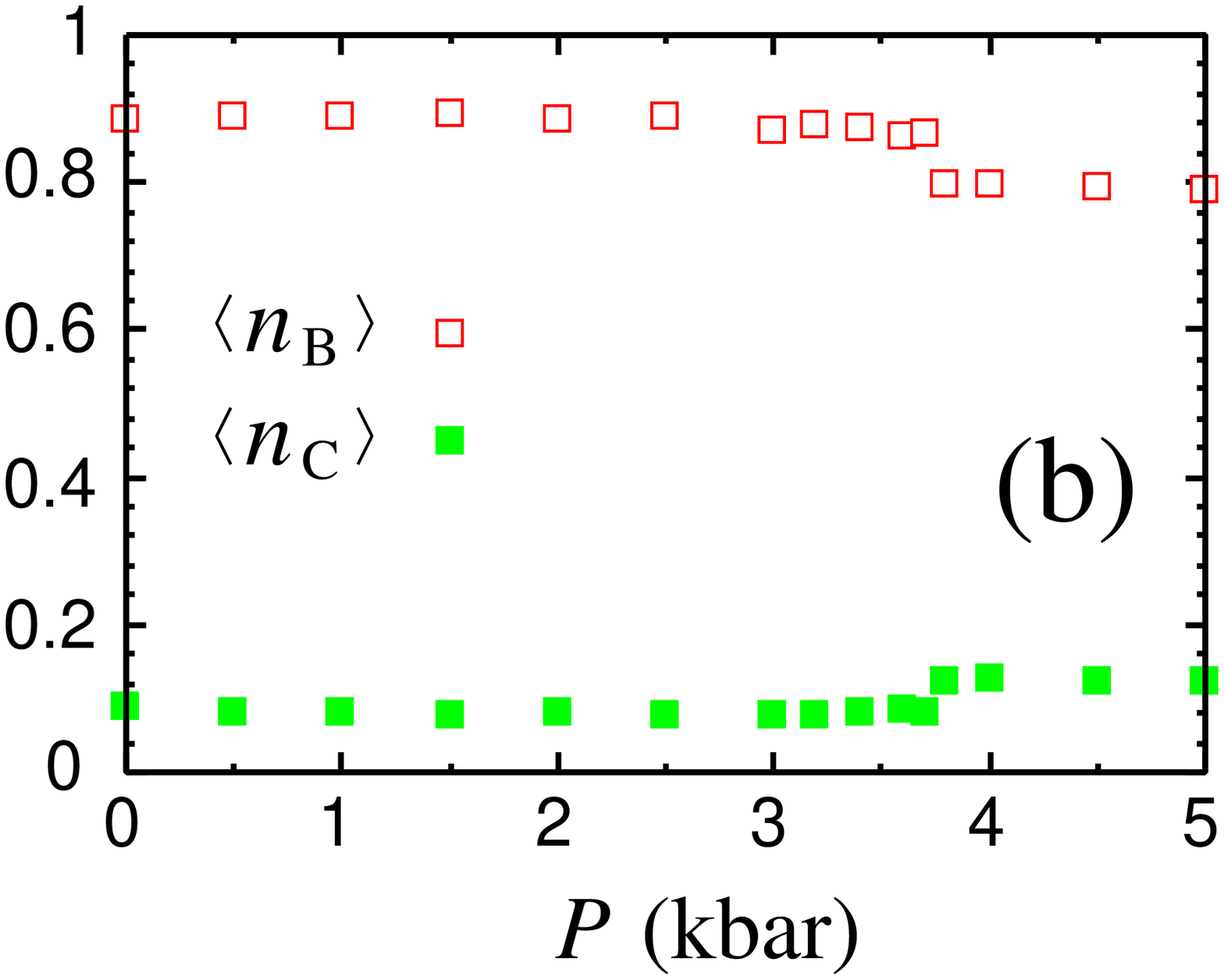}
\end{center}
\caption{(Color online) $P$ dependence of (a) $\langle n_{A}\rangle$ and $\langle n_{A^{\prime}}\rangle$, 
and (b) $\langle n_{B}\rangle$ and $\langle n_{C}\rangle$ with $U=0.7$, $V_a=0.35$, and $V_b=0.25$.}
\label{fig:fig3}
\end{figure}

\subsection{Ground-state properties under uniaxial pressure}
In the following, we show the fully optimized results with all the variational 
parameters. %Next, we show the results obtained with fully optimized wave functions. 
Note that the CO state is always nonmagnetic, which is in sharp contrast with the 
MFA\cite{Seo_JPSJ00,Kobayashi_JPSJ05,Tanaka_JPSJ08,Kobayashi_PRB11}. 
In Figs. \ref{fig:fig3}(a) and \ref{fig:fig3}(b), we plot $\langle n_{\alpha}\rangle$  
($=\langle n_{\alpha \uparrow}\rangle +\langle n_{\alpha \downarrow}\rangle$) 
as a function of the uniaxial pressure $P$ (kbar). 
When $P=0$ kbar, the ground state is the nonmagnetic CO. 
With increasing $P$, the charge disproportionation between the sites $A$ and 
$A^{\prime}$ gradually decreases. At $P\sim 3.8$ kbar, 
there is a first-order transition 
where $\langle n_{A}\rangle$ and $\langle n_{A^{\prime}}\rangle$ become equal, indicating a 
disappearance of 
the CO. For $P>3.8$ kbar, the inversion symmetry that is broken by the 
CO is recovered. 
$\langle n_{B}\rangle$ and $\langle n_{C}\rangle$ are not affected by $P$ except the 
transition point. They remain to be different through the transition since these sites are 
inequivalent in the crystal structure. The site $B$ ($C$) is 
charge-rich (charge-poor) even in the noninteracting case\cite{Kobayashi_JPSJ04}. 
\begin{figure}
\begin{center}
\includegraphics[height=4.2cm]{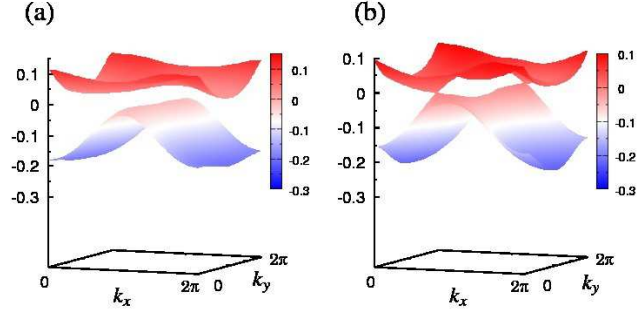}
\end{center}
\caption{(Color online) Energy dispersion of the lowest two bands for 
(a) $P=0$ kbar and (b) $P=5$ kbar. 
The other parameters are the same as those in Fig. \ref{fig:fig3}. The Fermi energy is taken as zero.}
\label{fig:fig4}
\end{figure}

In Figs. \ref{fig:fig4}(a) and \ref{fig:fig4}(b), we show the energy dispersion of the 
lowest two bands for $P=0$ and $5$ kbar, respectively. 
They are obtained by computing the eigenvalues of $H^{\rm MF}$ with sufficiently large system 
sizes, although the variational parameters are optimized with $N_u=8\times 8$. 
We confirmed that the results are qualitatively unchanged if the variational parameters with 
$N_u=6\times 6$ are employed. 
At $P=0$ kbar, there is a charge gap because of the existence of CO. 
For $P=5$ kbar, the energy dispersion indicates that a ZGS with two Dirac cones is realized. 
The Fermi level coincides with the contact points located at ${\bf k}^{\pm}_D=(\pm 0.70\pi,\pm 0.31\pi)$. 
The appearance of the Dirac cones by uniaxial pressure is consistent with the mean-field 
study\cite{Kobayashi_JPSJ05}. The present result indicates that the renormalization of 
$\Delta_{\alpha \sigma}$ 
by electron correlation does not affect the Dirac cones qualitatively. 

\begin{figure}
\begin{center}
\includegraphics[height=4.5cm]{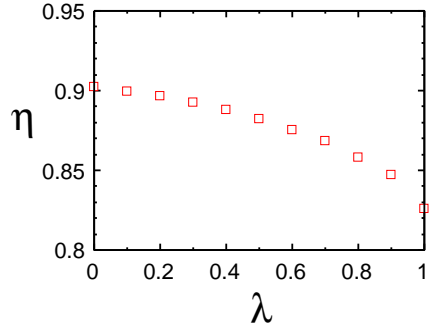}
\end{center}
\caption{$\lambda$ dependence of the tilting parameter $\eta$ at $P=5$ kbar.}
\label{fig:fig5}
\end{figure}

To investigate the correlation effects quantitatively, 
we consider the energy dispersion near the contact point. For $\alpha$-I$_3$, the 
Dirac cone has a tilting and its dispersion is written as\cite{Kobayashi_JPSJ07,Kobayashi_JPSJ08,Katayama_EPJ09,Suzumura_JPSJ14,Suzumura_JPSJ14_2}
\begin{equation}
\epsilon^{\pm}_{\bf k}=v_{x0}k_x+v_{y0}k_y\pm\sqrt{v^2_xk^2_x+v^2_yk^2_y},
\end{equation}
where $v_{x0}$ and $v_{y0}$ are the velocities for the tilting, whereas $v_{x}$ and 
$v_{y}$ are those of the Dirac cone. 
Here, ${\bf k}={\bf k}^{\pm}_D$ is taken as the origin. 
The degree of tilting is measured using the parameter $\eta$ ($0\leq \eta <1$), which is 
defined as\cite{Suzumura_JPSJ14,Suzumura_JPSJ14_2}  
\begin{equation}
\eta = \sqrt{(v_{x0}/v_x)^2+(v_{y0}/v_y)^2}. 
\end{equation}
We estimate $\eta$ by fitting the obtained energy bands to Eq. (6) near ${\bf k}={\bf k}^{\pm}_D$. 
The results are shown in Fig. \ref{fig:fig5} as a function of a parameter $\lambda$ that controls 
the strength of the interactions. We set $U=0.7\lambda$, $V_a=0.35\lambda$, $V_b=0.25\lambda$, 
and $P=5$ kbar, where the ZGS is the ground state for $0\leq \lambda \leq 1$. At $\lambda=0$, we have 
$\eta\sim 0.9$. Although $\eta$ gradually decreases with increasing $\lambda$, it is still 
larger than 0.8 at $\lambda=1$. These values are consistent with those obtained in 
previous mean-field studies\cite{Kobayashi_JPSJ08,Katayama_EPJ09}. 
For the values of $v_x$ and $v_y$, they slightly decrease as $\lambda$ increases. For example, 
we obtain $(v_x,v_y)=(0.0594,0.0537)$ at $P=0$ kbar and $(v_x,v_y)=(0.0527,0.0495)$ at 
$P=5$ kbar. Such a behavior is different from a recent renormalization group study with 
long-range 
Coulomb interactions\cite{Isobe_JPSJ12}, where $v_x$ and $v_y$ show an enhancement near 
the contact point. The decrease in these velocities may come from the short-range Coulomb 
interactions considered in this study. 

\begin{figure}
\begin{center}
\includegraphics[height=4.5cm]{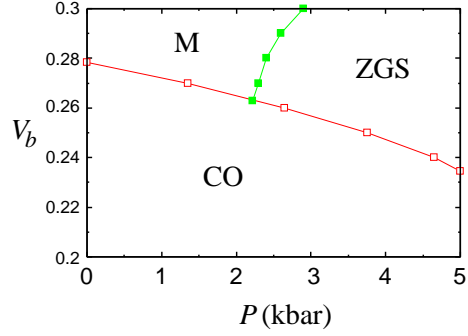}
\end{center}
\caption{Phase diagram on the $(P,V_b)$ plane for $U=0.7$ and $V_a=0.35$, where 
M indicates a metallic state that has electron and hole pockets in the Fermi surface.}
\label{fig:fig6}
\end{figure}

In Fig. \ref{fig:fig6}, we show the phase diagram in the $(P,V_b)$ plane for $U=0.7$ 
and $V_a=0.35$. The horizontal CO is dominant for small $V_b$. This is 
consistent with the MFA where the stabilities of various CO patterns are
compared\cite{Seo_JPSJ00,Tanaka_JPSJ08}. 
The realization of the horizontal CO comes from its CO pattern that can avoid the 
energy loss by $V_a$ as well as the band structure of $\alpha$-I$_3$. 
As $V_b$ increases, the horizontal CO is destabilized. This is considered to be 
due to charge frustration in a triangular lattice structure with $V_a\sim V_b$, which has been 
discussed for the case of $\theta$-(BEDT-TTF)$_2$X\cite{Seo_PRB05,Watanabe_JPSJ06,Nishimoto_PRB08}. 
Furthermore, the uniaxial pressure $P$ makes the CO unstable through the change in the band 
structure\cite{Kobayashi_JPSJ05}. 
In the large $P$ and $V_b$ region, the ZGS appears. On the other hand, 
in the region with small $P$ and large $V_b$, there is a metallic 
state that has electron Fermi surface and hole Fermi surface, which is basically the same 
as that in the noninteracting case at $P=0$ kbar\cite{Kobayashi_JPSJ05}. 

\subsection{Origin of the nonmagnetic CO}
\begin{figure}
\begin{center}
\includegraphics[height=4.5cm]{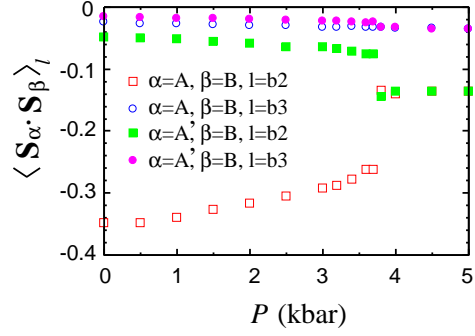}
\end{center}
\caption{(Color online) $P$ dependence of the nearest-neighbor 
spin-spin correlation functions $\langle {\bf S}_{\alpha}\cdot {\bf S}_{\beta}\rangle_l$ 
for $l=b2$ and $b3$. The other parameters are the same as those in Fig. \ref{fig:fig3}.}
\label{fig:fig7}
\end{figure}

To elucidate the origin of the nonmagnetic CO, we calculate the nearest-neighbor 
spin-spin correlation functions $\langle {\bf S}_{\alpha}\cdot {\bf S}_{\beta}\rangle_{l}$ 
as a function of $P$, where $\alpha$ and $\beta$ indicate a pair of sites in the unit 
cell and $l$ specifies the bond connecting these two sites. The results are shown in Fig. 
\ref{fig:fig7} for $l=b2$ and $b3$ (see Fig. \ref{fig:fig1}). In the CO state for $P<3.8$ kbar, 
the charge-rich stripe is found on the ABAB chain. On this charge-rich stripe, 
$|\langle {\bf S}_{A}\cdot {\bf S}_{B}\rangle_{b2}|$ is large while 
$|\langle {\bf S}_{A}\cdot {\bf S}_{B}\rangle_{b3}|$ is very small, which is reflected by 
the magnitude of $t_{b2}$ and $t_{b3}$ ($|t_{b2}|\sim 2|t_{b3}|$). This indicates that 
a spin-singlet is favored on the $b2$ bonds, which is markedly different from the results 
obtained in the MFA\cite{Seo_JPSJ00,Kobayashi_JPSJ05,Tanaka_JPSJ08,Kobayashi_PRB11}, 
but is consistent with the picture discussed from the NMR experiment\cite{Kawai_JPSJ09}. 
We consider that a one-dimensional spin-Peierls system is essentially realized 
because of the correlation effects that are incorporated in the projection operators. 
In fact, when a spin-singlet is formed on the b2 bonds, we expect 
$\langle {\bf S}_{A}\cdot {\bf S}_{B}\rangle_{b2}\sim
-\frac{3}{4}\langle n_{A} n_{B}\rangle_{b2}$\cite{Ogata_PRB90,Tanaka_JPSJ05} 
since a charge-charge correlation function for the b2 bonds, $\langle n_{A} n_{B}\rangle_{b2}$, 
is the probability of occurrence for two electrons 
to occupy the nearest-neighbor sites $A$ and $B$. Actually, at $P=0$ kbar, we have 
$-\frac{3}{4}\langle n_{A} n_{B}\rangle_{b2}=-0.406$ and 
$\langle {\bf S}_{A}\cdot {\bf S}_{B}\rangle_{b2}=-0.348$ so that 
the result is consistent with the formation of the spin-singlet. Experimentally, 
there is a structural distortion at the CO transition, which slightly increases $|t_{b2}|$ on the 
charge-rich stripe, whereas $t_{b3}$ is almost unchanged\cite{Kakiuchi_JPSJ07}. Although this may 
contribute to the spin-singlet formation, we consider that the lattice effect has only a minor role 
since the large bond alternation ($|t_{b2}|\sim 2|t_{b3}|$) already exists in the metallic phase.

For $P>3.8$ kbar, the CO disappears and the sites $A$ and $A^{\prime}$ become equivalent. 
An antiferromagnetic correlation exists on the two b2 bonds that connect a set of 
three sites $A$, $A^{\prime}$, and $B$. $\langle {\bf S}_{\alpha}\cdot {\bf S}_{\beta}\rangle_{l}$ 
are almost constant as a function of $P$ in the ZGS. 

On the other hand, $|\langle {\bf S}_{\alpha}\cdot {\bf S}_{\beta}\rangle_{b1}|$ (not shown) is small 
although $|t_{b1}|$ is large. This is because the site $C$ is always charge-poor 
so that the spin-spin correlation does not develop on the b1 bonds. 

\begin{figure}
\begin{center}
\includegraphics[height=5.0cm]{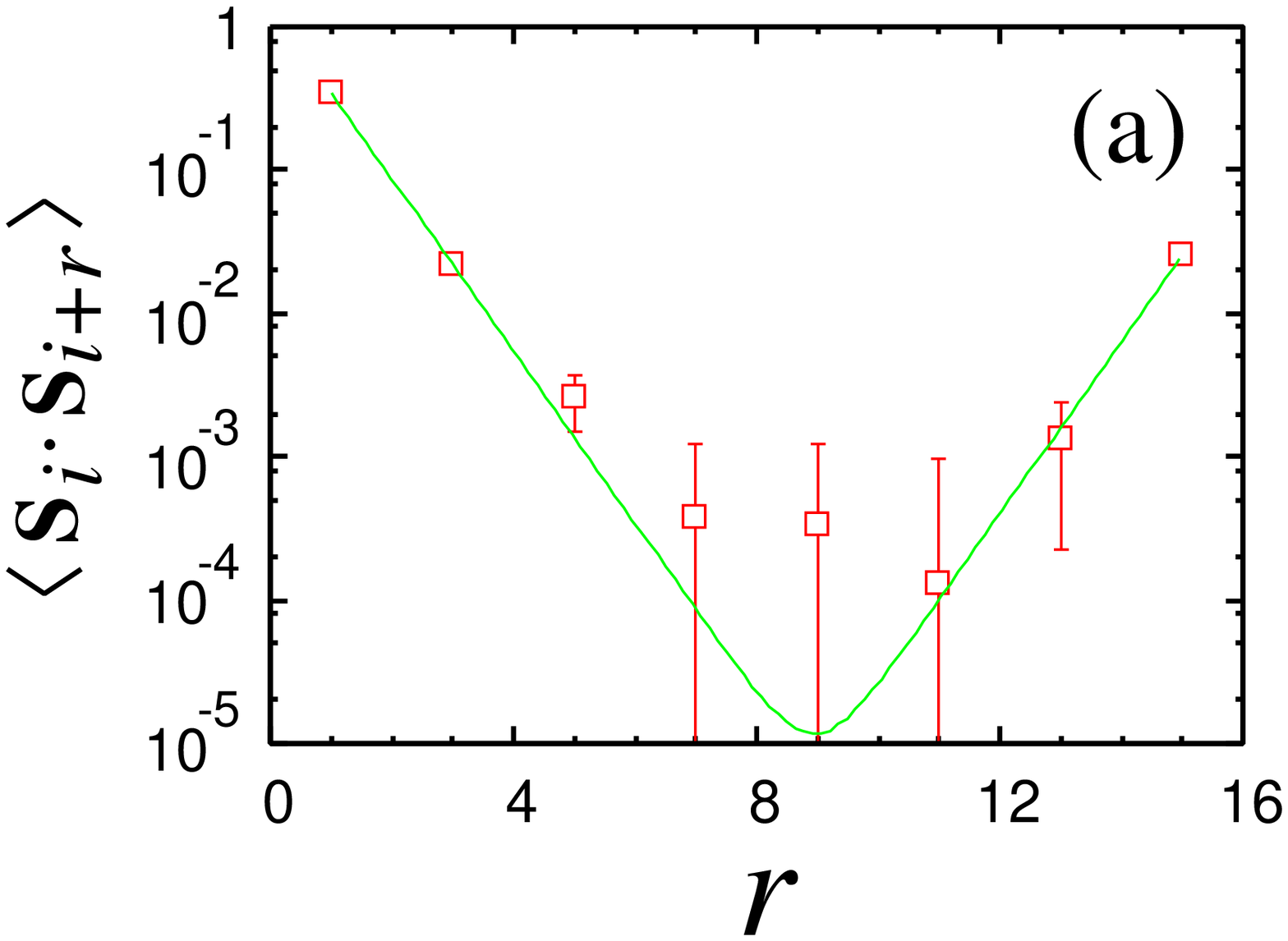}
\end{center}
\begin{center}
\includegraphics[height=4.5cm]{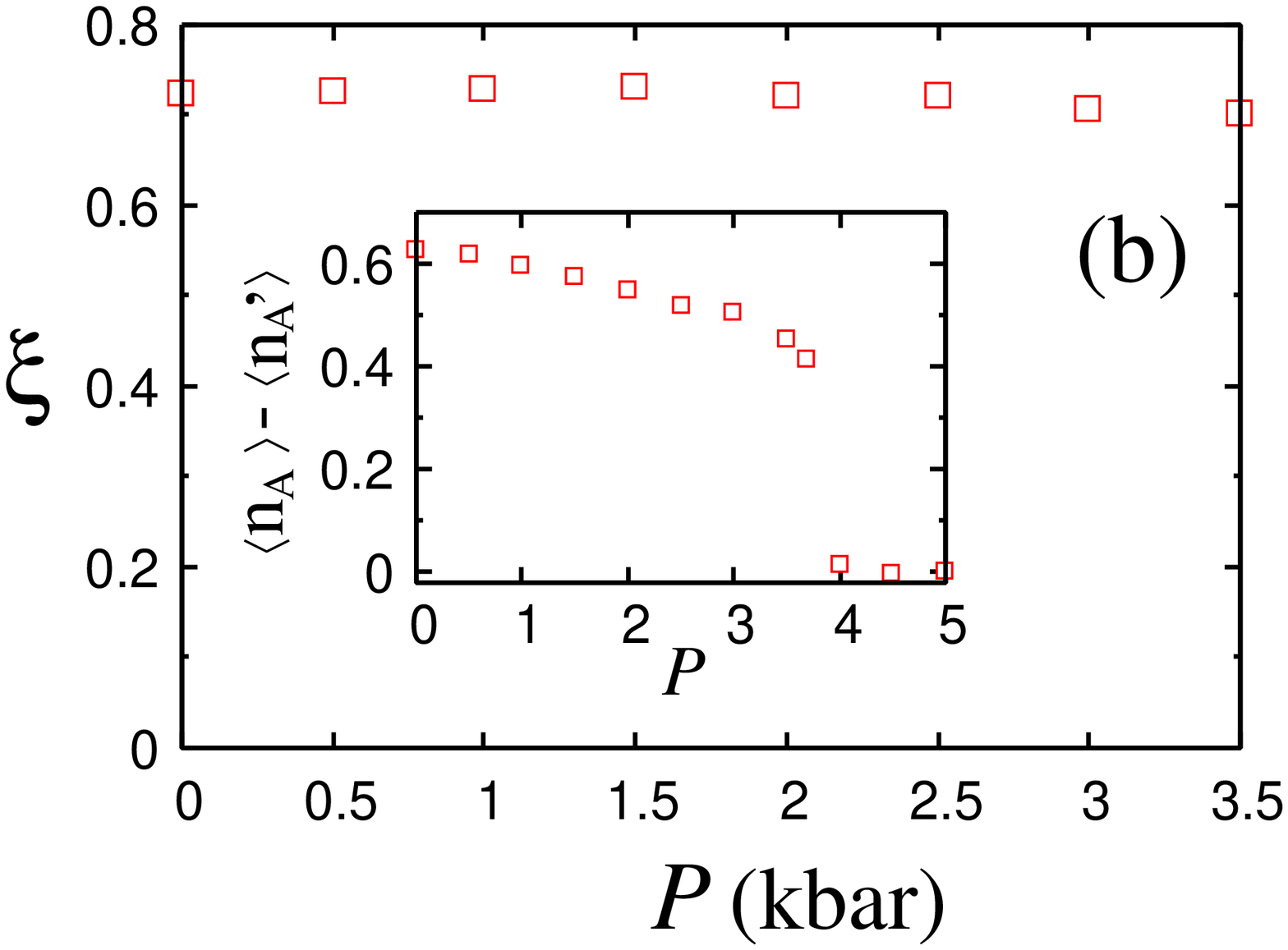}
\end{center}
\caption{(a) $r$ dependence of $\langle {\bf S}_{i}\cdot {\bf S}_{i+r}\rangle$ with 
$i=A$ at $P=0$ kbar and (b) the spin correlation length $\xi$ as a function of $P$. The inset shows the 
difference between the average charge densities on the sites $A$ and $A^{\prime}$. The other parameters 
are the same as those in Fig. \ref{fig:fig3}.}
\label{fig:fig8}
\end{figure}

\begin{figure}
\begin{center}
\includegraphics[height=4.5cm]{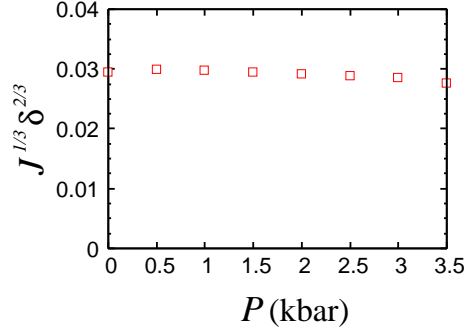}
\end{center}
\caption{$J^{1/3}\delta^{2/3}$ as a function of $P$.}
\label{fig:fig9}
\end{figure}
 
\subsection{Spin gap and CO}
Finally, we discuss the spin gap in the CO state. 
Figure \ref{fig:fig8}(a) shows the long-distance behavior of the spin-spin correlation 
function $\langle {\bf S}_{i}\cdot {\bf S}_{i+r}\rangle$ along the charge-rich stripe 
at $P=0$ kbar, where we set $i=A$ and the distance $r$ is measured along the stripe. 
Since the spin-gap behavior appears as an exponential decay of 
$\langle {\bf S}_{i}\cdot {\bf S}_{i+r}\rangle$, we fit the results as 
\begin{equation}
\langle {\bf S}_{i}\cdot {\bf S}_{i+r}\rangle=c_1e^{-r/\xi}+c_2e^{-(L-r)/\xi},
\end{equation}
where $c_1$, $c_2$, and the spin correlation length $\xi$ are fitting parameters. 
The fitting curve is also depicted in Fig. \ref{fig:fig8}(a). The result is in fact 
consistent with the existence of the spin gap. As shown in Fig. \ref{fig:fig8}(b), 
$\xi$ is almost constant as a function of $P$ indicating that the spin gap 
$\Delta_s$ has very weak $P$ dependence if we assume $\Delta_s\propto \xi^{-1}$. 
This behavior is in contrast to the suppression of the CO, which is manifested by a decrease 
in $\langle n_A\rangle -\langle n_{A^{\prime}}\rangle$, as shown in the 
inset. We note that the $P$ dependence of $\xi$ is also weaker than that of 
the nearest-neighbor spin-spin correlation function 
$\langle {\bf S}_{A}\cdot {\bf S}_{B}\rangle_{b2}$. Their 
pressure dependences are in general different from each other since $\xi$ is determined by 
the long-range correlation functions.

\section{Discussion}
Actually, recent transport and NMR measurements\cite{Liu_DT16,Liu_PRL16} have revealed that the 
spin gap is robust against hydrostatic pressure even in the region where $T_{\rm CO}$ and the 
charge gap are largely suppressed.

In order to understand the reason why $\Delta_s$ is roughly independent of $P$, we estimate 
$\Delta_s$ from another viewpoint. We regard the ABAB chain in the CO state as a one-dimensional 
spin system. Then, the effective spin exchange 
coupling on the b2 bonds is written as $J_{b2}\sim 4t_{b2}^2\langle n_An_B\rangle_{b2}/U$, 
whereas that on the b3 bonds is given by 
$J_{b3}\sim 4t_{b3}^2\langle n_An_B\rangle_{b3}/U$\cite{Ogata_PRB90,Tanaka_JPSJ05}. 
Therefore, this spin system can be regarded as a Heisenberg spin chain with alternating exchange 
couplings, $J_{b2}$ and $J_{b3}$. This model has a spin gap 
$\Delta_s\propto J^{1/3}\delta^{2/3}$, where 
$J=(J_{b2}+J_{b3})/2$ and $\delta =(J_{b2}-J_{b3})/2$\cite{Cross_PRB_79,Orignac_EPJB04}. 
In Fig. \ref{fig:fig9}, we plot $J^{1/3}\delta^{2/3}$ as a function of $P$. 
Apparently, the $P$ dependence of $\Delta_s$ is very weak and this estimation is consistent 
with the behavior of $\xi$ in Fig. \ref{fig:fig8}. The physical reason is as follows. 
When we increase $P$, $|t_{b2}|$ and $|t_{b3}|$ increase. However, $\langle n_An_B\rangle_{b2}$ and 
$\langle n_An_B\rangle_{b3}$ decrease with $P$ because of the suppression of the CO, the effect of 
which nearly cancels with the above enhancement. 

Finally, we remark on the difference between uniaxial and hydrostatic pressures. 
The latter has been used in several experiments such as NMR\cite{Liu_DT16,Liu_PRL16} and 
Raman\cite{Wojciechowski_PRB03} studies. Because 
the hydrostatic pressure essentially increases transfer integrals uniformly, its main effect 
is to increase the bandwidth\cite{Kondo_JPSJ09}. 
On the other hand, the change in transfer integrals is more anisotropic for the case of 
uniaxial pressure\cite{Kobayashi_JPSJ04}. 
Although the CO insulating state changes into the ZGS for both cases\cite{Tajima_JPSJ06}, 
there appear some differences in electronic properties: 
(i) $T_{\rm CO}$ has only weak pressure dependence under uniaxial
pressure compared with that under hydrostatic pressure\cite{Tajima_JPSJ06}. 
(ii) Theoretically, it has been suggested
that the effects of anion potentials may be important for the appearance of the ZGS under hydrostatic
pressure\cite{Suzumura_JPSJ13b}, 
which is different from the case of uniaxial pressure where it emerges even 
without the anion potentials. 
We expect that these differences do not have a significant role in the pressure dependence 
of spin and charge degrees of freedom obtained in this paper because of the following reasons. 
Experimentally, the resistivity under uniaxial pressure shows a substantial decrease as a function of 
pressure for $T<T_{\rm CO}$, indicating that the CO is
suppressed as in the case of hydrostatic pressure. In fact, at $T=0$, our results on 
$\langle n_A\rangle-\langle n_{A^{\prime}}\rangle$ show that it gradually decreases and vanishes 
discontinuously at a critical pressure, as shown in Fig. \ref{fig:fig8}(b). This is 
qualitatively consistent with the estimation from Raman experiments under hydrostatic 
pressure\cite{Wojciechowski_PRB03}. Therefore, we consider that point (i) does not affect 
our results essentially. The consistency between our results
and the Raman experiments also suggests that the anion potentials do not have a major role in 
the pressure dependence of the CO so that point (ii) will not be important at least in the CO 
state. 

For the spin gap, we consider that its weak pressure dependence is 
basically unaltered even when we take account of the hydrostatic pressure, because the hydrostatic pressure 
also induces 
the suppression of the CO together with the increase in the transfer integrals along the charge-rich 
stripe. As discussed in the present paper, these effects lead to the weak pressure dependence of $\Delta_s$ 
compared with the charge degrees of freedom. These considerations support the applicability of 
the present calculations to the experimental results under hydrostatic pressure.

\section{Summary}
In this paper, we investigate correlation effects on the ground-state properties in 
the organic conductor $\alpha$-I$_3$. 
In the CO state, we have shown that the antiferromagnetic spin order that is predicted by 
the MFA disappears owing to the correlation effects. The spins on the 
charge-rich stripe tend to form a spin-singlet on the bonds with large transfer integrals. 
Under uniaxial pressure, there is a transition from the nonmagnetic CO to the ZGS. 
We suggest that the uniaxial pressure dependence of the spin gap is very weak compared with 
that of the CO. These different behaviors in the spin and charge degrees of freedom are qualitatively 
consistent with experimental observations.

\begin{acknowledgments}
Y. T. is grateful to H. Watanabe for fruitful discussions. The authors would like to thank 
D. Liu, K. Miyagawa, and K. Kanoda for helpful discussions.
\end{acknowledgments}

\end{document}